\documentclass[12pt]{article}

\usepackage{amssymb,amsmath}

\usepackage{indentfirst}

\usepackage{multirow}

\usepackage{graphicx}
\DeclareGraphicsExtensions{.pdf,.png,.jpg}

 \catcode`\@=11 \@addtoreset{equation}{section}\catcode`\@=12
\newcommand{\R}{ {\mathbb R} }

\begin{document}


 \large{ \bf Examples of exact exponential cosmological solutions with three isotropic subspaces in the Einstein-Gauss-Bonnet gravity} 

 \vspace{0.3truecm}


   K. K. Ernazarov\footnote{e-mail: kubantai80@mail.ru}

\vspace{0.3truecm}

    Institute of Gravitation and Cosmology, \\
    Peoples' Friendship University of Russia (RUDN University), \\
    6 Miklukho-Maklaya Street,  Moscow, 117198, Russian Federation, \\ 
   


\begin{abstract}

We consider $(1+ 8)$-  and $(1+10)$-dimensional Einstein-Gauss-Bonnet models with the cosmological $\Lambda$-term. Some new examples of exact solutions with three constant Hubble-like parameters in this model are obtained, governed by three non-coinciding Hubble-like parameters: $H \neq 0$, $h_1$ and $h_2$, obeying  $m H + k_1 h_1 + k_2 h_2 \neq 0$, corresponding to factor spaces of dimensions $m\geqslant 3$, $ k_1 > 1$ and $ k_2\geqslant 1$. In this case, the multidimensional cosmological model deals with three factor spaces: the external 3-dimensional  "our"  world and internal subspaces with dimensions $ (m-3)$,  $k_1$  and  $k_2$.

\end{abstract}

  {\bf Keywords:} the multidimensional Einstein-Gauss-Bonnet gravity, Variation of G, Hubble-like parameters, cosmological term.

\section{Introduction}

Even before the Lovelock theory \cite{Lovelock}, a modified theory of gravity has been developed. As in the Einstein equation in the Lovelock theory the field equations are represented in the form of the second order \cite{Einstein}. Lovelock's modified theory differs from Einstein's theory in that a higher order of curvature is included in the equation. Under this condition, the Lagrangian of the field is defined as:
\begin{equation}
  S = \sum\limits_{s=0}^{[\frac{d-1}{2}]} \alpha_s L_s ,
 \label{1.1}
\end{equation}

where  $\alpha_s$ and $L_s$ are the s-order Lovelock parameter and Lagrangian, respectively.

Lovelock's gravity  \cite{Lovelock} is based on the metric and curvature tensor and is one of  generalized theories of gravity. The Lagrangian includes the scalar curvature, the cosmological term, and new corrections for each odd spacetime dimension above four. The first correction is called the Gauss-Bonnet term and is presented in the following form:
\begin{equation}
  G = R_{abcd} R^{abcd} - 4R_{ab}R^{ab} + R^2.
 \label{1.2}
\end{equation}

This term defines multidimensional gravity, i.e., gravity in the case of $D>4$ . It can be noted here that in the four-dimensional case these corrections are either topological or identically zero, and Einstein's theory in four dimensions follows from the new modified theory of gravity.

After the advent of the general relativity (GR), it is believed that the Einstein equations are the most effective theory describing physical phenomena from large scales to small scales \cite{Einstein}. However, recent research data argue that there are some limitations in explaining some of the new observations: based on general relativity, we cannot discover the nature of dark matter \cite{Young} and dark energy \cite{Weinberg}-\cite{Perlmutter}. From this we can conclude that to create a new theory of gravity, which describes the dynamics of the universe and the nature of its dark side, the need for modern theoretical physics.

Corrections to Einstein's $\mathrm{4D}$ gravity have different motivations. There is such a motivation that by increasing the dimension from four space-times, one can construct a new theory, which follows from it the four-dimensional theory of Einstein (see, for example, \cite{Arkani-H} - \cite{Arkani-H-2}), others on scalar tensor couplings, that could give cosmologically observable effects \cite{Wett}, \cite{Wett-2} or violate the equivalence principle \cite{Damour}. In another motivation for the new alternative theory of gravity, a correction is established by adding quadratic combinations of the Riemann tensor (see, for example, \cite{Parry}), such as $R^2$ and $R_{\mu\rho}R^{\mu\rho}$ , i.e. higher-order modification field equations. Modification theories have been used in the past to model inflation \cite{Starobinsky}, \cite{Capoz-Amendola} and dark energy \cite{Capoz}. The modified gravity, which eliminates the need for dark energy and which seems to be stable is considered \cite{Nojiri:2003}. The terms with positive powers of the curvature support the inflationary epoch while the terms with negative powers of the curvature serve as effective dark energy, supporting current cosmic acceleration. The equivalent scalar-tensor gravity may be compatible with the simplest solar system experiments. It is possible to construct a unique quadratic combination of the Riemann curvature tensor that, when added to the usual Einstein-Hilbert action, does not increase the differential order of the equations of motion \cite{Lovelock}, \cite{Zumino}. The correction is obtained from the Gauss-Bonnet theorem concerning the Euler characteristic of two-dimensional surfaces \cite{Spivak} and is therefore called the Gauss-Bonnet term. In a multidimensional theory of gravity, there is no significant reason to omit it (other than to make it more complicated). In multidimensional gravity, mostly in the context of brane worlds, such theories have been shown to have surprising properties. Without trying to be completed, we can refer to the physics of black holes (e.g. \cite{Boulware}, \cite{Dufaux}), gravitational instabilities (e.g. \cite{Charm} ), and the dynamics of worlds on branes of codimension one and higher (e.g. \cite{Cho} - \cite{Charmousis}). General model of multidimensional $\mathrm{R}^2$-gravity including Riemann tensor square term (non-zero $\mathnormal{c}$ case) is considered in \cite{Nojiri:2001ae}.  The article is reported that, the number of brane-worlds in such model is constructed (mainly in five dimensions) and their properties are discussed. Thermodynamics of $\mathrm{S}$-$\mathrm{AdS}$ $\mathrm{BH}$ (with boundary) is presented when perturbation on $\mathnormal{c}$  is used. The entropy, free energy and energy are calculated. For non-zero $\mathnormal{c}$  the entropy (energy) is not proportional to the area (mass). The equation of motion of brane in $\mathrm{BH}$ background is presented as $\mathrm{FRW}$ equation. Using dual $\mathrm{CFT}$ description it is shown that dual field theory is not conformal one when $\mathnormal{c}$  is not zero. Thus, by adding the Gauss-Bonnet term, it is possible to construct a modified theory - multidimensional gravity. This term  in four dimensions is reduced to a complete divergence and, as such, has no dynamic significance. However, in the case of the scalar-tensor theory (see, for example, \cite{Damour2} for a detailed and general discussion), the Gauss-Bonnet term is coupled to the scalar sector, and four-dimensional gravity is modified (see also \cite{Campbell} for the case of the Lorentz and Chern-Simons terms ). Therefore, the Gauss-Bonnet term is included in the most general second-order scalar-tensor theory. Such a connection appears if we consider, for example, the case when the scalar field is a moduli field, which comes from the toroidal (i.e., flat) Kaluza-Klein (KK) compactification of the $(4 + N)$-dimensional theory. But such reasoning about the uniqueness of the Gauss-Bonnet term is valid only at the classical level.
Another motivation is related to low-energy efficient string actions, such as in heterotic string theory. In string coupling, the Gauss-Bonnet term is included as the leading order and the unique (before this order) ghostless $\alpha'$ correction to the Einstein-Hilbert term \cite{Metsaev} (see, for example, \cite{Antoni} for cosmology and models of the world on a brane arising in as a result of such actions). Since the related Gauss-Bonnet term is included in the scalar-tensor theory, an important question arises: how can current observations limit such models? It has already been shown in the literature that at the background level a coupled Gauss-Bonnet term allows for a viable cosmology \cite{Nojiri}.

\section{The cosmological model}

The action of the model reads
\begin{equation}
  S =  \int_{M} d^{D}z \sqrt{|g|} \{ \alpha_1 (R[g] - 2 \Lambda) +
              \alpha_2 {\cal L}_2[g] \},
 \label{2.0}
\end{equation}
where $g = g_{MN} dz^{M} \otimes dz^{N}$ is the metric defined on
the manifold $M$, ${\dim M} = D$, $|g| = |\det (g_{MN})|$, $\Lambda$ is
the cosmological term, $R[g]$ is scalar curvature,
$${\cal L}_2[g] = R_{MNPQ} R^{MNPQ} - 4 R_{MN} R^{MN} +R^2$$
is the standard Gauss-Bonnet term and  $\alpha_1$, $\alpha_2$ are
nonzero constants.

We consider the manifold
\begin{equation}
   M = \R  \times   M_1 \times \ldots \times M_n 
   \label{2.1}
\end{equation}
with the metric
\begin{equation}
   g= - d t \otimes d t  +
      \sum_{i=1}^{n} B_i e^{2v^i t} dy^i \otimes dy^i,
  \label{2.2}
\end{equation}
where   $B_i > 0$ are arbitrary constants, $i = 1, \dots, n$, and
$M_1, \dots,  M_n$  are one-dimensional manifolds (either $\R$ or $S^1$)
and $n > 3$.

The equations of motion for the action (\ref{2.0}) 
give us the set of  polynomial equations \cite{ErIvKob-16}
\begin{eqnarray}
  E = G_{ij} v^i v^j + 2 \Lambda
  - \alpha   G_{ijkl} v^i v^j v^k v^l = 0,  \label{2.3} \\
   Y_i =  \left[ 2   G_{ij} v^j
    - \frac{4}{3} \alpha  G_{ijkl}  v^j v^k v^l \right] \sum_{i=1}^n v^i 
    - \frac{2}{3}   G_{ij} v^i v^j  +  \frac{8}{3} \Lambda = 0,
   \label{2.4}
\end{eqnarray}
$i = 1,\ldots, n$, where  $\alpha = \alpha_2/\alpha_1$. Here
\begin{equation}
G_{ij} = \delta_{ij} -1, \qquad   G_{ijkl}  = G_{ij} G_{ik} G_{il} G_{jk} G_{jl} G_{kl}
\label{2.4G}
\end{equation}
are, respectively, the components of two  metrics on  $\R^{n}$ \cite{Ivas-2010, Ivas-2010-2}. 
The first one is a 2-metric and the second one is a Finslerian 4-metric.
For $n > 3$ we get a set of forth-order polynomial  equations.

We note that for $\Lambda =0$ and $n > 3$ the set of equations (\ref{2.3}) 
and (\ref{2.4}) has an isotropic solution $v^1 = \cdots = v^n = H$ only 
if $\alpha  < 0$ \cite{Ivas-2010, Ivas-2010-2}.
This solution was generalized in \cite{Chirkov} to the case $\Lambda \neq 0$.

It was shown in \cite{Ivas-2010, Ivas-2010-2} that there are no more than
three different  numbers among  $v^1,\dots ,v^n$ when $\Lambda =0$. This is valid also
for  $\Lambda \neq 0$ if $\sum_{i = 1}^{n} v^i \neq 0$  \cite{Ivas-2016}.


Here we consider a class of solutions to the set of equations (\ref{2.3}), 
(\ref{2.4}) of the following form:
\begin{equation}
  \label{3.1}
   v =(\overbrace{H, \ldots, H}^{m}, 
   \overbrace{h_1, \ldots, h_1}^{k_1}, \overbrace{h_2, \ldots, h_2}^{k_2}),
\end{equation}
where $H$ is the Hubble-like parameter corresponding to an $m$-dimensional factor space,  $h_1$ is the Hubble-like parameter 
corresponding to an $k_1$-dimensional factor space with $k_1 > 1$ and $h_2$ is the Hubble-like parameter corresponding to an $k_2$-dimensional factor space with $k_2 \geqslant 1$. We split the $m$-dimensional  factor space for $m > 3$ into the  product of two subspaces of dimensions $3$ and $m-3$, respectively. The first one is identified with ``our'' $3d$ space while the second one is considered as a subspace of $(m-3 + k_1 + k_2)$-dimensional internal space.

We consider the ansatz (\ref{3.1}) with three Hubble-like parameters $H$, $h_1$ and $h_2$  which obey the following restrictions:
   
\begin{equation}
     H \neq h_1, \quad  H \neq h_2, \quad  h_1 \neq h_2, 
   \label{3.3DD}
   \end{equation}
and 
\begin{equation}
     S_1 = m H + k_1 h_1 + k_2 h_2 \neq 0. 
   \label{3.3S1}
   \end{equation}

Among the exact solutions to the set of equations (\ref{2.3}) and (\ref{2.4}) with $ k_2 = 1$ 
are of interest, i.e. when 

\begin{equation}
  \label{3.2}
   v =(\overbrace{H, \ldots, H}^{m}, 
   \overbrace{h_1, \ldots, h_1}^{k_1}, h_2 ).
\end{equation}

There  solutions exist for certain value of  $ \lambda = \Lambda\alpha$ ($k_2 = 1$):

\begin{equation}
  \label{3.3}
   \lambda  = \frac{1}{8}\cdot\frac{(m+1)(mk_1 + 1) - 4mk_1 + (m - k_1)^2}{(m-1)(k_1 - 1)(m + k_1 -2)}.
\end{equation}

For this value of $\lambda$  Hubble-like parameters are defined as following:

\begin{equation}
  \label{3.4}
   H = \sqrt{\frac{(k_1 - 1)}{2\alpha(m-1)(m + k_1 - 2)}},
\end{equation}

\begin{equation}
  \label{3.5}
  h_1 = -\frac{m - 1}{k_1 - 1}H,
\end{equation}

\begin{equation}
  \label{3.6}
  h_2 \in \mathbb{R}.
\end{equation}

Among stable solutions, obeying $H >0$, $k_1 > 1$, 
$k_2 > 1$,  and the key stability condition \cite{Ivas-2016}
 \begin{equation}
      S_1  = m H + k_1 h_1 + k_2 h_2 > 0, 
    \label{3.3S2}
    \end{equation}
 solutions with zero variation of the gravitational constant $G$ were of great interest.
(We note that for $k_1 = 1$ or  $k_2 = 1$ the analysis of stability from Ref. \cite{Ivas-2016}  does not work. )
 Stable solutions that describe the exponential expansion in the $3d$ - subspace with the Hubble parameter $H > 0$ and with zero variation of the effective gravitational constant $G$ were found in \cite{ErIv-2017}. In this case, the condition of zero variation of the effective gravitational constant $G$ should be  also added to the system of polynomial equations:

\begin{equation}
  \label{3.7}
  (m-3)H + k_1h_1 + k_2h_2 = 0.  
\end{equation}

\section{Analysis of solutions for $(1+8)$-  and $(1+10)$-dimensional models}

\subsection{$(1+8)$-dimensional model}

Among the solutions in $(1+8)$-dimensional model can be identified as following four stable solutions with $ H > 0$:

In $[3,3,2]$-splitting solutions to the set of equations (\ref{2.3}), (\ref{2.4}) of the following form:

\begin{equation}
  \label{3.8}
   v = (H, H, H, h_1, h_1, h_1, h_2, h_2  ),
\end{equation}
 
 where $H$ is the Hubble-like parameter corresponding to an 3-dimensional factor space,  $h_1$  is the Hubble-like parameter corresponding to an 3-dimensional factor space and $h_2$  is the Hubble-like parameter corresponding to an 2-dimensional factor space.

1) In this case, the following two stable solutions among eight can be found for $ H>0$ and $\alpha > 0$:

1A)

\begin{equation}
  \label{3.9}
   h_2 = \frac{1}{\sqrt{15\alpha}}\sqrt{1 - 2\sqrt{60\lambda - 11}} > 0 ,
\end{equation}

\begin{equation}
  \label{3.10}
   h_1 = -\frac{1}{4\sqrt{15\alpha}}\left( \sqrt{1 - 2\sqrt{60\lambda - 11}} + 5\sqrt{1 + \frac{2}{5}\sqrt{60\lambda - 11}}\right) < 0 ,
\end{equation}

\begin{equation}
  \label{3.11}
   H = \frac{1}{4\sqrt{15\alpha}}\left(5\sqrt{1 + \frac{2}{5}\sqrt{60\lambda - 11}} - \sqrt{1 - 2\sqrt{60\lambda - 11}}\right) >  0 ,
\end{equation}

\begin{equation}
  \label{3.11AN}
  S_1 = \frac{1}{2\sqrt{15\alpha}}\sqrt{1 - 2\sqrt{60\lambda - 11}} > 0 .
\end{equation}

The interval of $\lambda$, which occurs for the stable solutions with $ H > 0$, $ h_1 < 0$  and $h_2 > 0$  is

\begin{equation}
\label{3.11BN}
     \frac{11}{60} <  \lambda < \frac{3}{16}.
\end{equation}

1B)

\begin{equation}
  \label{3.12}
   h_2 = \frac{1}{\sqrt{15\alpha}}\sqrt{1 + 2\sqrt{60\lambda - 11}} > 0 ,
\end{equation}

\begin{equation}
  \label{3.13}
   h_1 = -\frac{1}{4\sqrt{15\alpha}}\left( \sqrt{1 + 2\sqrt{60\lambda - 11}} + 5\sqrt{1 - \frac{2}{5}\sqrt{60\lambda - 11}}\right) < 0 ,
\end{equation}

\begin{equation}
  \label{3.14}
   H = \frac{1}{4\sqrt{15\alpha}}\left(5\sqrt{1 - \frac{2}{5}\sqrt{60\lambda - 11}} - \sqrt{1 + 2\sqrt{60\lambda - 11}}\right) >  0 ,
\end{equation}

\begin{equation}
  \label{3.14A}
  S_1 = \frac{1}{2\sqrt{15\alpha}}\sqrt{1 + 2\sqrt{60\lambda - 11}} > 0 .
\end{equation}

The interval of $\lambda$, which occurs for the stable solutions with $ H > 0$, $ h_1 < 0$  and $h_2 > 0$  is

\begin{equation}
\label{3.14B}
     \frac{11}{60} < \lambda < \frac{1}{4}.
\end{equation}

The  solution 1B) was presented (in fact) earlier in Ref. \cite{ErIv-2020-GC}. 
Here we eliminate the mistake in upper bound on $\lambda$ in relation (3.21) from \cite{ErIv-2020-GC}. \\ 

2) In $[m, k_1, 1]$- splitting solutions to the set of equations  (\ref{2.3}), (\ref{2.4}) are following: \\

\begin{tabular}{lllll}
 \bf 1.  & \bf m = 3,    & $ \bf k_1 = 4$, & $ \bf k_2 = 1$, & $ \bf \lambda = \frac{11}{60}$.   \\
\multicolumn{5}{l}{$H = \frac{1}{2}\sqrt{\frac{3}{5\alpha}}$, $h_1 = -\frac{1}{\sqrt{15\alpha}}$, $h_2 \in \mathbb{R}$.}   \\
\\
 \bf 2. & \bf m = 4,   & $ \bf k_1 = 3$, & $ \bf k_2 = 1$, & $ \bf \lambda = \frac{11}{60}$.    \\
\multicolumn{5}{l}{$H = \frac{1}{\sqrt{15\alpha}}$, $h_1 = - \frac{1}{2}\sqrt{\frac{3}{5\alpha}}$, $h_2 \in \mathbb{R}$.}   \\
\\
\bf  3.  & \bf  m = 5,   & $ \bf k_1 = 2$, & $ \bf k_2 = 1$, & $ \bf \lambda = \frac{23}{80}$.    \\
\multicolumn{5}{l}{$H = \frac{1}{2\sqrt{10\alpha}}$, $h_1 = - \sqrt{\frac{2}{5\alpha}}$, $h_2 \in \mathbb{R}$.}   \\
\end{tabular} 
\\
\\

3) Using formulas (3.17), (3.20) and (3.25) from \cite{ErIv-2017}, we can find stable solutions at the $ \lambda = \Lambda \alpha = \frac{213}{980}$ with $[3, 3, 2]$- spitting in dimension $(1+ 8)$ with zero variation of the effective gravitational constant G. These solutions are expressed as follows:

\begin{displaymath}
    H = \frac{1}{2\sqrt{35\alpha}},
\end{displaymath}

\begin{displaymath}
    h_1 = -\frac{2}{\sqrt{35\alpha}},
\end{displaymath}

\begin{displaymath}
   h_2 = \frac{3}{\sqrt{35\alpha}}.
\end{displaymath}

\subsection{$(1+10)$-dimensional model}

Among the solutions in $(1+ 10)$-dimensional model can be identified as following four stable solutions with $ H > 0$:

In $[4,4,2]$-splitting solutions to the set of equations (\ref{2.3}), (\ref{2.4}) of the following form:

\begin{equation}
  \label{3.8}
   v = (H, H, H, H, h_1, h_1, h_1, h_1, h_2, h_2  )
\end{equation}
 
 where H is the Hubble-like parameter corresponding to an 4-dimensional factor space,  $h_1$  is the Hubble-like parameter corresponding to an 4-dimensional factor space and $h_2$  is the Hubble-like parameter corresponding to an 2-dimensional factor space.

1) In this case, the following two stable solutions among eight can be found for $ H>0$ and $\alpha > 0$:

1A)

\begin{equation}
  \label{3.9A}
   h_2 = \frac{1}{28}\sqrt{\frac{42}{\alpha}}\sqrt{1 - \sqrt{336\lambda - 55}} > 0 ,
\end{equation}

\begin{equation}
  \label{3.10A}
   h_1 = -\frac{1}{4\sqrt{42\alpha}}\left( \sqrt{1 - \sqrt{336\lambda - 55}} + 7\sqrt{1 + \frac{1}{7}\sqrt{336\lambda - 55}}\right) < 0 ,
\end{equation}

\begin{equation}
  \label{3.11A}
  H = \frac{1}{4\sqrt{42\alpha}}\left( 7\sqrt{1 + \frac{1}{7}\sqrt{336\lambda - 55}} -  \sqrt{1 - \sqrt{336\lambda - 55}} \right) > 0,
\end{equation}

\begin{equation}
  \label{3.14AF}
  S_1 = \frac{1}{\sqrt{42\alpha}}\sqrt{1 - \sqrt{336\lambda - 55}} > 0 .
\end{equation}

The interval of $\lambda$, which occurs for the stable solutions with $ H > 0$, $ h_1 < 0$  and $h_2 > 0$  is

\begin{equation}
\label{3.14AR}
     \frac{55}{336} < \lambda < \frac{1}{6}.
\end{equation}

1B)

\begin{equation}
  \label{3.12A}
   h_2 = \frac{1}{28}\sqrt{\frac{42}{\alpha}}\sqrt{1 + \sqrt{336\lambda - 55}} > 0 ,
\end{equation}

\begin{equation}
  \label{3.13A}
   h_1 = -\frac{1}{4\sqrt{42\alpha}}\left( \sqrt{1 + \sqrt{336\lambda - 55}} + 7\sqrt{1 - \frac{1}{7}\sqrt{336\lambda - 55}}\right) < 0 ,
\end{equation}

\begin{equation}
  \label{3.14A}
  H = \frac{1}{4\sqrt{42\alpha}}\left( 7\sqrt{1 - \frac{1}{7}\sqrt{336\lambda - 55}} -  \sqrt{1 + \sqrt{336\lambda - 55}} \right) > 0,
\end{equation}

\begin{equation}
  \label{3.15FM}
  S_1 = \frac{1}{\sqrt{42\alpha}}\sqrt{1 + \sqrt{336\lambda - 55}} > 0 .
\end{equation}

The interval of $\lambda$, which occurs for the stable solutions with $ H > 0$, $ h_1 < 0$  and $h_2 > 0$  is

\begin{equation}
\label{3.16FM}
     \frac{55}{336} < \lambda < \frac{13}{48}.
\end{equation}

2) In $[ m,  k_1, 1]$- splitting solutions to the set of equations (\ref{2.3}), (\ref{2.4})  are following:\\

\begin{tabular}{lllll}
  \bf 1.  & \bf m = 3,   & $\bf k_1 = 6$, & $ \bf k_2 = 1$,   & $ \bf \lambda = \frac{27}{140}$.  \\
\multicolumn{5}{l}{$H = \frac{1}{2}\sqrt{\frac{5}{7\alpha}}$, $h_1 = -\frac{1}{\sqrt{35\alpha}}$, $h_2 \in \mathbb{R}$.}   \\
\\
  \bf 2.  & \bf m = 4,   & $\bf k_1 = 5$, & $ \bf k_2 = 1$,   & $ \bf \lambda = \frac{55}{336}$.  \\
\multicolumn{5}{l}{$H = \frac{1}{\sqrt{21\alpha}}$, $h_1 = - \frac{1}{4}\sqrt{\frac{6}{7\alpha}}$, $h_2 \in \mathbb{R}$.}   \\
\\
 \bf 3.  & \bf m = 5,   & $\bf k_1 = 4$, & $ \bf k_2 = 1$,   & $ \bf \lambda = \frac{55}{336}$.  \\   
\multicolumn{5}{l}{$H = \frac{1}{4}\sqrt{\frac{6}{7\alpha}}$, $h_1 = -\sqrt{\frac{2}{21\alpha}}$, $h_2 \in \mathbb{R}.$}   \\
\\
\bf 4.  & \bf m = 6,   & $\bf k_1 = 3$, & $ \bf k_2 = 1$,   & $ \bf \lambda = \frac{27}{140}$.  \\  
\multicolumn{5}{l}{$H = \frac{1}{\sqrt{35\alpha}}$, $h_1 = - \frac{1}{2}\sqrt{\frac{5}{7\alpha}}$, $h_2 \in \mathbb{R}.$}  \\
\\
\bf 5.  & \bf m = 7,   & $\bf k_1 = 2$, & $ \bf k_2 = 1$,   & $ \bf \lambda = \frac{13}{42}$.  \\  
\multicolumn{5}{l}{$H = \frac{1}{2}\sqrt{\frac{1}{21\alpha}}$, $h_1 = -\sqrt{\frac{3}{7\alpha}}$, $h_2 \in \mathbb{R}.$}   \\
\\ 
\end{tabular}

	Stable solutions at the well-defined values of  $\lambda = \Lambda\alpha$  with $[3, 5, 2]$-, $[4, 4, 2]$- and $[5, 3, 2]$- splitting  in dimension $(1+10)$ with zero variation of the effective gravitational constant G are expressed as follows:\\

\begin{tabular}{lllll}
\bf 1.  & \bf m = 3,   & $\bf k_1 = 5$, & $ \bf k_2 = 2$,   & $ \bf \lambda = \frac{991}{4732}$.  \\ 
\multicolumn{5}{l}{$H = \frac{3}{2}\sqrt{\frac{1}{91\alpha}}$, $h_1 = -\frac{2}{\sqrt{91\alpha}}$, $h_2 = 5\sqrt{\frac{1}{91\alpha}}$.}   \\
\\
\bf 2.  & \bf m = 4,   & $\bf k_1 = 4$, & $ \bf k_2 = 2$,   & $ \bf \lambda = \frac{269}{1344}$.  \\ 
\multicolumn{5}{l}{$H = \frac{1}{2\sqrt{21\alpha}}$, $h_1 = - \frac{5}{4\sqrt{21\alpha}}$, $h_2 = \frac{9}{4\sqrt{21\alpha}}$.}   \\
\\
\bf 3.  & \bf m = 5,   & $\bf k_1 = 3$, & $ \bf k_2 = 2$,   & $ \bf \lambda = \frac{589}{2800}$.  \\ 
\multicolumn{5}{l}{$H = \frac{1}{2\sqrt{70\alpha}}$, $h_1 = - 3\sqrt{\frac{1}{70\alpha}}$, $h_2 = 2\sqrt{\frac{2}{35\alpha}}$.}   \\
  \\
\end{tabular}

\section{Conclusions}

We have considered the $(1 + 8)$-  and  $(1+10)$-dimensional Einstein-Gauss-Bonnet (EGB) models with a $\Lambda$-term. By using the ansatz with diagonal cosmological metrics, we have found for different splitting and certain $\lambda = \alpha\Lambda$ a class of explicit solutions with three Hubble-like parameters $H > 0$ , $h_1$, and $ h_2$  corresponding to submanifolds of dimensions $ m \geqslant 3$, $k_1>1$, and $k_2 \geqslant 1$  respectively. The obtained solutions are exact.  They are stable for $ k_2 > 1$. As we know, stability plays a predominant role in exact solutions of a set of equations. All the exact solutions obtained in this paper have been verified through the results of previously published papers \cite{ErIv-2020}, \cite{Ivas-2016}. \\

 {\bf Acknowledgments}

The publication has been prepared with the support of the  ''RUDN University Program 5-100''. The reported study was funded by RFBR, project number 19-02-00346. Author is grateful to  V.D. Ivashchuk for discussions.


\end{document}